\documentclass[prl, aps, twocolumn, groupedaddress, superscriptaddress]{revtex4}
\usepackage{epsfig}
\usepackage{slashed}
\usepackage{graphicx}
\usepackage[usenames, dvipsnames]{color}
\usepackage{graphics}
\usepackage{hyperref}
\usepackage{bm}
\usepackage{subfigure}

\begin{document}

\title{Driven collective instabilities in magneto-optical traps: a fluid-dynamical approach}

\author{Hugo Ter\c{c}as}
\email{htercas@cfif.ist.utl.pt}
\affiliation{CFIF, Instituto Superior T\'{e}cnico, Av. Rovisco Pais 1, 1049-001 Lisboa, Portugal}
\author{J. Tito Mendon\c ca}
\affiliation{CFIF, Instituto Superior T\'{e}cnico, Av. Rovisco Pais 1, 1049-001 Lisboa, Portugal}
\affiliation{IPFN, Instituto Superior T\'{e}cnico, Av. Rovisco Pais 1, 1049-001 Lisboa, Portugal}
\author{Robin Kaiser}
\affiliation{Institut Non Lin\'eaire de Nice, UMR 6618,  1361 Route des Lucioles, F-06560 Valbonne, France}

\begin{abstract}
We present a theoretical model to describe an instability mechanism in ultra-cold gases, where long-range interactions are taken into account. Focusing on the nonlinear coupling between the collective (plasma-like) and the center-of-mass modes, we show that the resulting dynamics is governed by a parametric equation of the generalized Mathieu type and compute the corresponding stability chart. We apply our model to typical ranges of magneto-optical traps (MOT) parameters and find a good agreement with previous experimental observations.   
\end{abstract}


\maketitle


The celebrated method of magneto-optical cooling of atoms \cite{cohen} allowed in recent years the exploitation of new and very interesting phenomena in atomic physics, and in particular the creation of Bose-Einstein condensates \cite{BEC}. Cold and ultracold gases produced in a magneto-optical trap (MOT) provide an unique medium to explore the connections between different areas of physics. Due to absorption and re-scattering of light \cite{walker,dalibard,sesko}, collective phenomena of cold gases confined in a MOT became recently an issue of major interest. Under such conditions, the atomic gas in a MOT can be regarded as a fluid with a tunable effective charge, which opens a new area of plasma physics \cite{steane, mendonca, pruvost}. A very interesting manifestation of the latter is the work on the Coulomb explosions of molasses, performed by Pruvost {\it et al} \cite{pruvost}. Moreover, in a recent work, we have proposed the existence of hybrid waves and Tonks-Dattner oscillations in the confined gas of ultra-cold atoms, very similar but not identical to those of a plasma medium \cite{mendonca}. \par

It is a well known fact that cold atomic gases confined in MOTs sustain various types of instabilities. One example is found in the work realized by Kim {\it et al.} \cite{kim}, where a parametric instability is excited by an intensity modulated laser beam. This instability is based on a nonlinear response of the individual atoms to one of the control parameters of the MOT. On the other side, spontaneous instabilities arising for large MOTs have been described in \cite{david, stefano, hennequin}, where the feedback of retroreflected laser beams of the MOT can induce stochastic or deterministic chaos for a large optical thickness of the MOT. These instabilities have been investigated by looking at the motion of the center of mass of the cloud. A further type of spontaneous instabilities, due to the large spatial extend of the MOT has been described in \cite{labeyrie}. Here, the instability finds its origin at the edge of the MOT where the Zeeman shift produces a local change of the detuning seen by the atoms with a negative friction for atoms beyond a certain horizon.\par

In the present work, we theoretically explore the possible existence of a new instability process, due to the coupling between two collective oscillations in a MOT:  the ''plasma'' and the center-of-mass (c.o.m.) modes. We use the formalism developed in our recent work \cite{mendonca}, where we have shown that plasma-hybrid waves can propagate in the atomic cloud and have established their dispersion relation. These hybrid waves exhibit features that are common to both electron-plasma and acoustic waves \cite{cheng, parker, guerra}. Here we start with a set of fluid equations describing the dynamics of the atomic medium, where we have retained the main features of the cooling and trapping forces. We then show that these plasma-like oscillations can couple with the center-of-mass of the cloud, giving place to the a possible parametric instability mechanism. By identifying the relevant parameters and establishing the stability criteria, we are able to show that this mechanism could lead to an instability threshold in the same parameter range as the one observed by Labeyrie {\it et al.} \cite{labeyrie}. New instability regimes are also predicted.


Following our previous work \cite{mendonca}, the collective behavior in the trap is governed by a set of fluid equations

\begin{equation}
\frac{\partial n}{\partial t}+\bm{\nabla}\cdot(n\mathbf{v})=0,\label{eq:2.6a}
\end{equation}

\begin{equation}
\frac{\partial\mathbf{v}}{\partial t}+(\mathbf{v}\cdot\bm{\nabla})\mathbf{v}=\frac{\mathbf{F}}{M},\label{eq:2.6b}
\end{equation}

\begin{equation}
\bm{\nabla}\cdot\mathbf{F_c}=Qn
\label{eq:2.6c},
\end{equation}
where $n$ and $\mathbf{v}$ represent the gas density and velocity, respectively, and $M$ is the atomic mass. The total force is given by $\mathbf{F}=\mathbf{F}_{c}+\mathbf{F}_{t}$, where $\mathbf{F}_{c}$ is the collective force and $Q=(\sigma_{R}-\sigma_{L})\sigma_{L}I_{0}/c$ represents the square of the effective electric charge \cite{mendonca, pruvost}, with $c$  being the speed of light, $I_{0}$ the total intensity of the six laser beams and $\sigma_{R}$ and $\sigma_{L}$ representing the emission and absorption cross sections \cite{walker}. The term $\mathbf{F}_{t}=-\bm{\nabla}U$ stands fot the trapping force. We consider a low intensity Doppler regime MOT, where the trapping potential is given by

\begin{equation}
U(\mathbf{r})= \frac{1}{2}\kappa r^2 \; , \quad \kappa=\alpha \frac{\mu_{B}}{\hbar k} \nabla B.
\label{single}
\end{equation}
Here, $\mu_{B}$ represents the Bohr magneton, $\kappa=\kappa(\delta, I_{0}/I_{s})$ is the spring constant, $\alpha=\alpha(\delta, I_{0}/I_{s})$ is the friction coefficient, $\delta$ is the laser detuning, $I_{s}$ is the atomic saturation intensity and $\nabla B=\vert \bm{\nabla}B\vert$ represents the magnetic field gradient \cite{mendonca}. In such a picture, the gas confined in a MOT can be regarded as an one-component plasma, where the effective plasma frequency is given by 
\begin{equation}
\omega_P=\sqrt{Q n_0 / M}. 
\label{plasma}
\end{equation}
It is assumed that the equilibrium density of the gas $n_{0} \equiv n_{0}(r)$ depends only the radial coordinate $r$ and angular distortions of the cloud will be ignored. For simplicity, we will use a water-bag model, where we assume that the density is approximately constant inside the cloud. \par
We linearize Eqs. (\ref{eq:2.6a})-(\ref{eq:2.6c}) by separating each relevant physical quantity into its equilibrium and perturbation components, such that

\begin{equation}
n=n_0+n_1+n_{2}, \quad \mathbf{v}=\mathbf{v}_1+\mathbf{v}_{2}, \quad\mathbf{F}=\mathbf{F}_{1}+\mathbf{F}_{2}.
\end{equation}
The subscripts $1$ and $2$ label two different perturbations, corresponding to two distinct collective modes in the MOT. This procedure allows one to include any pair of modes and could be extended to more than two modes, thus broadening the range of applicability of the model. In this work, we consider that $1$ and $2$ label the perturbations due to center-of-mass (c.o.m.) and plasma oscillations. Accordingly, $\mathbf{F}_{1}=\mathbf{F}_{t}$ is the restoring force in the c.o.m. and $\mathbf{F}_{2}=\mathbf{F}_{c}$ represents the collective force responsible for the plasma oscillations. For moderate oscillation amplitudes, the center-of-mass velocity $\mathbf{v}_1$ is described by a harmonic oscillation

\begin{equation}
\mathbf{v}_{1}(t)=\mathbf{u}_{1}\sin(\omega_{CM}t+\phi)
\label{single2},
\end{equation}
where $\omega_{CM}$ (associated to $\mathbf{F}_{1}$ - for the sake of generality, the details on the calculation of $\omega_{CM}$ will be discussed later on) represents the c.o.m. mode frequency, $\mathbf{u}_{1}$ its amplitude and $\phi$ is an arbitrary phase. We will consider that the c.o.m. oscillates as a rigid body, which means that the perturbation $n_{1}$ can be neglected ($n_1 \ll n_2\ll n_0$). Combining these approximations with the above fluid equations, one easily obtains

\begin{eqnarray}
\label{mathieu}
\frac{\partial^2 n_{2}}{\partial t^2}&+&n_0 \frac{\partial}{\partial t}\bm{\nabla}\cdot\mathbf{v}_{2}+\nonumber\\
&+&\mathbf{u}_1\cdot\bm{\nabla}\left[ \frac{\partial  n_{2}}{\partial t}\sin(\omega_{CM}t+\phi)+\right.\\
&+& \left. n_{2} \omega_{CM}\cos(\omega_{CM}t+\phi)\right]=0,\nonumber
\end{eqnarray}
Noticing that $n_{0}\partial_{t}\bm{\nabla}\cdot \mathbf{v}_{2}=\omega_{P}^2n_{2}$, it follows

\begin{eqnarray}
\label{mathieu2}
\frac{\partial^2 n_{2}}{\partial t^2}&+&\omega_P^2 n_{2}+\nonumber\\
&+&\mathbf{u}_1\cdot\bm{\nabla} \left[\frac{\partial  n_{2}}{\partial t}\sin(\omega_{CM}t+\phi)+\right.\\
&+&\left. \omega_{CM}n_2\cos(\omega_{CM}t+\phi) \right]=0.\nonumber
\end{eqnarray}
The third term of this expression describes the nonlinear coupling between the two collective modes mentioned above. This is one of the main results of the paper. Notice that the plasma mode (\ref{plasma}) could be replaced by any of those in the hierarchy $\omega_{n,l}$ associated to the Tonks-Dattner (TD) resonances, the standing waves inside the cloud which hierarchy provides a large number of modes with frequencies very close to each other \cite{mendonca}. The reason for this choice is because $\omega_{P}$ is the lowest mode in the TD series. \par
\begin{figure}
\includegraphics[scale=0.55]{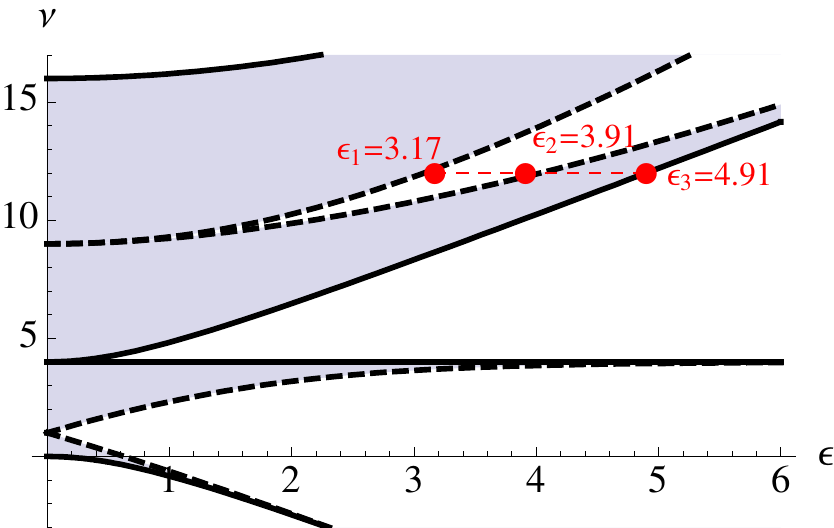}
\caption{Stability chart of Eq.(\ref{eq:2.12}), obtained for a $n=5$ Hill determinant. The full lines represent $\pi$-period solutions and the dashed lines represent the $2\pi$-period ones. The shadowed (light) regions correspond to stable (unstable) solutions. The red dots represent the marginal values $\epsilon_{1}$, $\epsilon_{2}$ and $\epsilon_{3}$ discussed in the text.}
\label{fig1}
\end{figure}

Making use of the \textit{ansatz} $ n_{2}(\mathbf{r},t)= B_{2}(\mathbf{r})A_{2}(t)$, Eq. (\ref{mathieu2}) yields
 
\begin{equation}
\frac{d^2 A_{2}}{d \tau^2}+\left[\nu +2\epsilon\cos (2\tau)\right]A_{2}+\epsilon\sin (2\tau)\frac{d A_{2}}{d \tau}=0,
\label{eq:2.12}
\end{equation}
where $\xi=2\mathbf{u}_1\cdot\bm{\nabla}\ln  B_{2}$ is a free nonlinear coupling parameter and describes the amplitude of the center-of-mass oscillation. Here, $2\tau=\omega_{CM}t+\phi$, $\nu=4\omega_{P}^2/\omega_{CM}^2$ and $\epsilon=\xi/\omega_{CM}$ are dimensionless variables. Equation (\ref{eq:2.12}) describes the dynamics of a parametrically excited system and belongs to the family of Hill equations. It is formally similar to the Mathieu equation, which is well-known in the literature for containing unstable solutions. According to the standard Floquet theory \cite{nayfeh}, the solutions to Eq. (\ref{eq:2.12}) can be expanded into Fourier series
\begin{equation}
A_{2}(\tau)=\sum_{n=-\infty}^{+\infty}a_{n}e^{(\gamma+in)\tau},
\end{equation}
where $\gamma$ is the characteristic exponent \cite{nayfeh}. Plugging it into Eq. (\ref{eq:2.12}), one obtains a system of equations for the coefficients $a_{n}$. The non-trivial solution is obtained by computing the roots of the so-called infinite Hill determinant, i.e.,

\begin{eqnarray}
\label{eq:Hill}
\mbox{det}\left(\Delta_{j,k}(n)\right)&=&\det\left(\left[\nu+(\gamma+2jn)^2\right]\delta_{j,k}\right. \nonumber\\
&+&\epsilon\left[1-\frac{i}{2}\left(\gamma+2jn\right)\right]\delta_{j,k+1}\\
&+&\epsilon\left.\left[1+\frac{i}{2}\left(\gamma+2jn\right)\right]\delta_{j,k-1}\right)=0,\nonumber
\end{eqnarray}
where $\delta_{j,k}$ is the Kroenecker delta. In order to carry out calculations, only the truncated system $n=-N,...,N$ is considered. This approximation results on the Ince-Strutt diagram \cite{strutt}. In Fig. (\ref{fig1}) we plot the stability chart of Eq. (\ref{eq:2.12}) in the $\nu-\epsilon$ plan, truncated at $N=5$, where the $\pi$ and $2\pi$-periodic marginal curves correspond to $\gamma=0$ and $\gamma=i$, respectively. The marginal curves $\nu(\epsilon)$ separate the different regions of stability (check Ref. \cite{nayfeh} for further details on the Floquet theory).

\begin{figure}
\subfigure[]{
\includegraphics[scale=0.47]{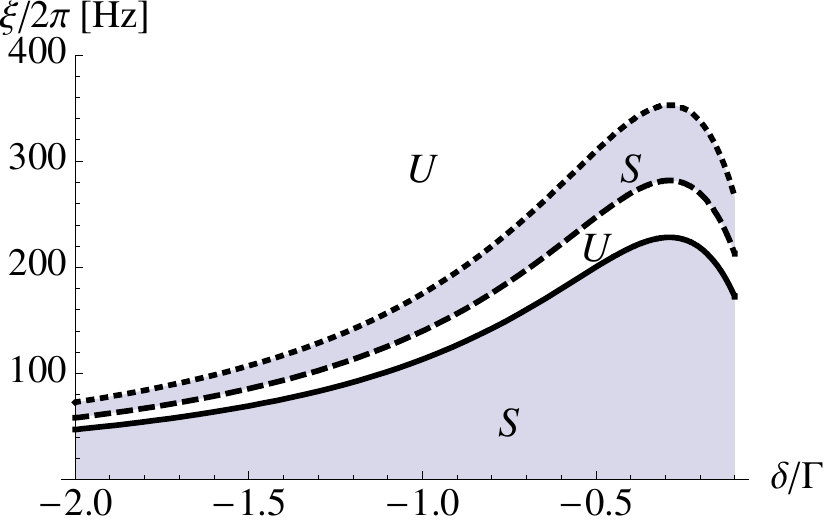}}
\subfigure[]{
\includegraphics[scale=0.5]{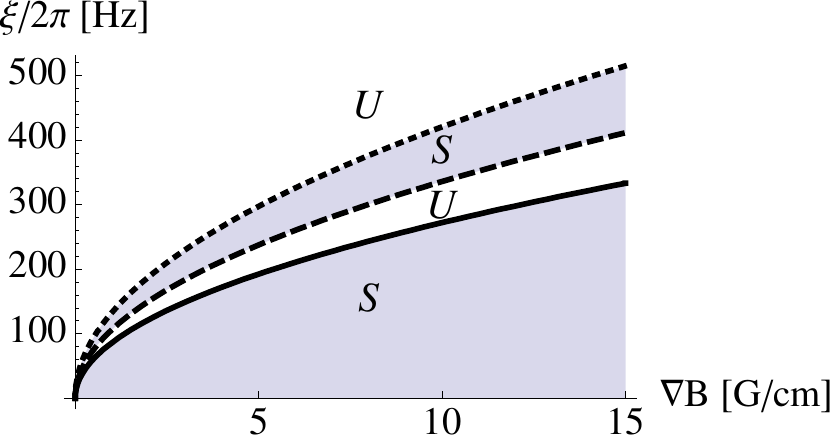}}
\caption{(Color online) Marginal curves $\xi_{i}=\xi(\epsilon_{i})$ $(i=1,2,3)$ as function of: a) the detuning $\delta$, at $\nabla B=3$ G/cm and b) the magnetic field gradient $\nabla B$, at $\delta=-0.75\Gamma$. In both cases,  $I_{0}/I_{s}=0.3$ and $\xi_{1}$ (solid line), $\xi_{2}$ (dashed line) and $\xi_{3}$ (dotted line) are the marginal curves discussed in the text. The shadowed areas are stable.}
\label{fig2}
\end{figure}

For comparison with experiments, it is necessary to express the frequencies $\omega_{P}$ and $\omega_{CM}$ in terms of real-life parameters. The balance between the trapping and collective forces is given by the hydrodynamical equilibrium condition $D/D_{t}\mathbf{v}\equiv(\partial/\partial_{t} +\mathbf{v}\cdot\bm{\nabla})\mathbf{v}=0$, which simply corresponds to set 

\begin{equation}
\mathbf{F}_{t}+\mathbf{F}_{c}=0,
\end{equation} 
where $\mathbf{F}_{t}=-\bm{\nabla}U=-\kappa \mathbf{r}$. Taking the divergence of the later, we obtain the following relation
\begin{equation}
\kappa=\frac{(\sigma_{R}-\sigma_{L})\sigma_{L}I_{0}}{3c}=\frac{Qn_{0}}{3},
\label{limit}
\end{equation}
which simply corresponds to the expression for the compression limit of the MOT \cite{sesko}. This condition establishes a relation between the plasma (\ref{plasma}) and the c.o.m oscillation frequencies: $\omega_{CM}=\omega_{t}=\omega_{P}/\sqrt{3}$, where $\omega_{t}=\sqrt{\kappa/M}$ is the trapping frequency. The identity $\omega_{CM}=\omega_{t}$ is equivalent of setting $\mathbf{F}_{1}=\mathbf{F}_{t}$ and it is simply a consequence of the Kohn's theorem \cite{kohn}, which basically states that interactions do not affect the c.o.m. motion. The equality $\omega_{CM}=\omega_{P}/\sqrt{3}$ is often recognized in unscreened Coulomb systems \cite{seidl, mie}, as it describes the so-called Mie resonance. The analogy between MOTs and Coulomb systems follows from the fact that the effect of the trap is replaced by the background ionic density in the later. For typical experimental conditions of a cold $^{85}$Rb gas, operating at $\delta=-1.5\Gamma$, where $\Gamma\approx 6$ MHz is the natural lifetime of the atomic transition, with a magnetic field gradient $\nabla B=5$ G/cm and $I_{0}/I_{s}\approx 0.3$, the corresponding plasma frequency is $\omega_{P}/2\pi\approx 120$ Hz. As a consequence of the condition (\ref{limit}), we set the value $\nu=12$ in Eq. (\ref{eq:2.12}) and, correspondingly, the instability occurs provided that

\begin{equation}
\epsilon_{1}<\epsilon<\epsilon_{2}\quad \mbox{and} \quad \epsilon >\epsilon_{3},
\label{marginal}
\end{equation}
where $\epsilon_{1}=3.17$, $\epsilon_{2}=3.91$ and $\epsilon_{3}=4.91$ (see Fig. (\ref{fig1})). Using the definition of $\epsilon$, and restricting the discussion to the low saturation limit, the marginal curves $\xi=\xi(\epsilon_{i})$, with $i=(1,2,3)$, representing the critical values of the nonlinear coupling parameter, are simply given by

\begin{equation}
\left(\frac{\xi}{\epsilon_{i}}\right)^2=\omega_{CM}^2=8\frac{k_{L}\mu_{B} \nabla B}{M}\frac{I_{0}}{I_{s}}\frac{\vert\delta\vert/\Gamma}{\left(1+4\delta^2/\Gamma^2\right)^2},
\label{parameters}
\end{equation}
where $k_{L}$ is the laser wave vector. In Fig. (\ref{fig2}) we plot the marginal curves $\xi_{1}$, $\xi_{2}$ and $\xi_{3}$ against the relevant MOT parameters. Below the marginal curve $\xi_{1}$, the oscillations are stable. However, for $\xi_{1}<\xi<\xi_{2}$, the system undergoes large amplitude density oscillations and the becomes unstable. The oscillations become stable between the marginal curves $\xi_{2}$ and $\xi_{3}$. Finally for $\xi>\xi_{3}$ we can observe that the oscillations are again unstable.\par

\begin{figure}
\includegraphics[scale=0.65]{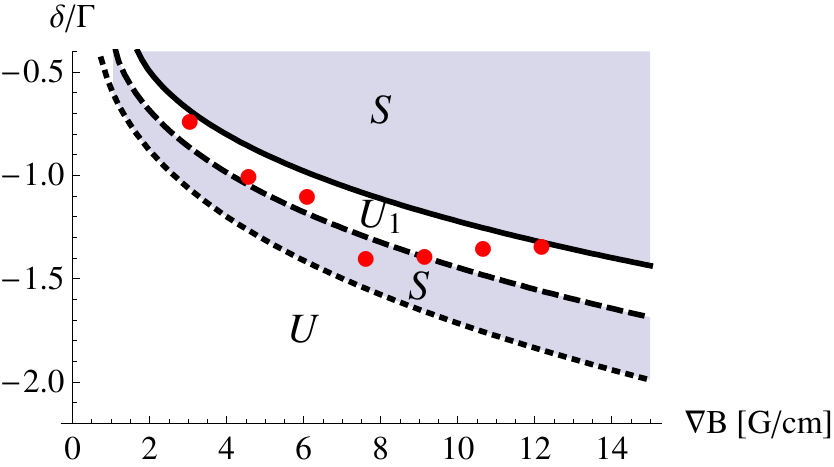}
\caption{(Color online) Iso-$\xi$ curves obtained for $I_{0}/I_{s}=0.3$ and $\xi/2\pi\approx280$ Hz, for the different marginal values of $\epsilon$: $\epsilon_{1}$ (full line), $\epsilon_{2}$ (dashed line) and $\epsilon_{3}$ (dotted line). The shadowed (light) regions are stable (unstable). The red dots represent the experimental threshold in Ref. \cite{labeyrie}.}
\label{fig3}
\end{figure}

This way of coupling different modes of the system, added to an external drive (or sufficient power at the correct frequency in the noise spectrum) and therefore controlling the value $\xi$, could lead to an instability based on mechanisms similar to the ones used in \cite{kim}. Applying this idea to a typical range of ($\delta$, $\nabla B$) in Eq. (\ref{parameters}), it is possible to build a stability chart similar to that presented in Fig. (\ref{fig3}). We note that a stability diagram similar to the one observed by Labeyrie et al. \cite{labeyrie} can be obtained for specific values of driving parameters, corresponding here to $\xi=280$ Hz. In the present paper, however, the basic ingredient for the instability is different from that used in \cite{labeyrie}, what therefore suggests it is possible to distinguish these two mechanisms: in the later, the instability is due to a change of sign in the friction coefficient at the edge of the cloud \cite{pohl}, where the driving motion takes place. Instead, we here suggest that an instability in the same parameter range can be obtained by coupling two (or more) collective modes, where the energy is transfered from one to the other, as an example of a density-wave mixing phenomenon. Our model also predicts different stability zones, which does not happen in the model of Ref. \cite{labeyrie}. We believe that the main difference is the fact that our model does not depend on the size of the cloud, which is consistent with the fluid approximation. As a consequence, we do not require the edge to contribute to the dynamics of the system. We also argue that the coupling between the c.o.m and the different TD modes may occur in real experiments (as they are very close to $\omega_{P}$) and it should be responsible for the broadening of the intermediate instability zone $U_{1}$ in Fig. \ref{fig3}, which better agrees with the experimental observations of Ref. \cite{labeyrie}. This is because for $\nu>12$ the intermediate instability zone $\epsilon_{1}<\epsilon<\epsilon_{2}$ is larger, as one can directly observe in Fig. (\ref{fig1}).\par
A short discussion of the validity range of the present model is in order. We expect the present results to be valid only valid in the low-intensity Doppler regime, where the dynamics of the single atom MOT remains essentially linear. The existence of nonlinear corrections is obvious from the above fluid equations, which are not at all related with the nonlinearity in the basic forces, but are only rooted in the collective behavior of the MOT. \par


In this work, we have highlighted several aspects of the fluid description of the collective behavior of a cold atomic gas in a magneto-optical trap. We have established the governing equation for the nonlinearly coupled dynamics of the center-of-mass and the plasma oscillations of the cloud, which we have shown to be the root of an instability mechanism. It was also shown that the condition of dynamical equilibrium provides a relation between the frequencies of the center-of-mass and the plasma oscillations, $\omega_{CM}=\omega_{P}/\sqrt{3}$. Our model yields an instability threshold which can be close to the one observed in \cite{labeyrie}, even though mechanisms not included in this work have been put forward to explain these self-sustained instabilities.
In addition, we predict the existence of new stability regions, compatible with the same coupling parameter, thus giving a new insight to the problem and proposing new areas of experimental research. We concluded that our model can be used to describe, at least qualitatively, a family of instabilities of the collective oscillations in a MOT.\par
 One of the authors (H. Ter\c{c}as) would like to thank Dr. Paulo Oliveira, from the Electrical Engineering Department of Instituto Superior T\'{e}cnico, for helpful discussions and advice in the preparation of a preliminary version of this manuscript. This work was partly supported by Funda\c{c}\~{a}o para a Ci\^encia e Tecnologia (FCT-Portugal) through the grant number SFRH/BD/37452/2007.

\end{document}